\documentclass[reprint, superscriptaddress, amsmath, amssymb, aps]{revtex4-2}
\usepackage[letterpaper,margin=0.74in]{geometry}

\usepackage{xcolor}
\usepackage{braket}
\usepackage{subcaption}
\usepackage[nolist, nohyperlinks]{acronym}
\usepackage{amsmath, amsthm, amssymb, amsbsy, mathtools, mathalpha, bm}
\usepackage[normalem]{ulem}
\usepackage{cancel}
\usepackage{siunitx}
\usepackage{graphicx}
\usepackage{hyperref}
\usepackage{orcidlink}

\usepackage{natbib}
\newcommand{\coloredcite}[2][blue]{\hypersetup{citecolor=#1}\cite{#2}\hypersetup{citecolor=green}}

\usepackage{caption}
\makeatletter
\renewcommand*\env@matrix[1][\arraystretch]{%
  \edef\arraystretch{#1}%
  \hskip -\arraycolsep
  \let\@ifnextchar\new@ifnextchar
  \array{*\c@MaxMatrixCols c}}
\makeatother

\newcommand{\vect}[1]{\boldsymbol{#1}}

\begin{document}

\title{Likelihood for a Network of Gravitational-Wave Detectors with Correlated Noise}

\author{Francesco Cireddu\,\orcidlink{0009-0002-7074-4278}}
\affiliation{Leuven Gravity Institute, KU Leuven, Celestijnenlaan 200D box 2415, 3001 Leuven, Belgium}
\affiliation{Department of Physics and Astronomy, Laboratory for Semiconductor Physics, KU Leuven, B-3001 Leuven, Belgium}
\affiliation{Dipartimento di Fisica “E. Fermi”, Università di Pisa, I-56127 Pisa, Italy}
\author{Milan Wils\,\orcidlink{0000-0002-1544-7193}}
\affiliation{Leuven Gravity Institute, KU Leuven, Celestijnenlaan 200D box 2415, 3001 Leuven, Belgium}
\affiliation{Department of Physics and Astronomy, Laboratory for Semiconductor Physics, KU Leuven, B-3001 Leuven, Belgium}
\author{Isaac C. F. Wong\,\orcidlink{0000-0003-2166-0027}}
\affiliation{Leuven Gravity Institute, KU Leuven, Celestijnenlaan 200D box 2415, 3001 Leuven, Belgium}
\affiliation{Department of Physics, The Chinese University of Hong Kong, Shatin, N.T., Hong Kong}
\affiliation{KU Leuven, Department of Electrical Engineering (ESAT), STADIUS Center for Dynamical Systems, Signal Processing and Data Analytics, B-3001 Leuven, Belgium}
\author{Peter T. H. Pang\,\orcidlink{0000-0001-7041-3239}}
\affiliation{Nikhef, Science Park 105, 1098 XG Amsterdam, The Netherlands}
\affiliation{Institute for Gravitational and Subatomic Physics (GRASP), Utrecht University, Princetonplein 1, 3584 CC Utrecht, The Netherlands}
\author{Tjonnie G. F. Li\,\orcidlink{0000-0003-4297-7365}}
\affiliation{Leuven Gravity Institute, KU Leuven, Celestijnenlaan 200D box 2415, 3001 Leuven, Belgium}
\affiliation{Department of Physics and Astronomy, Laboratory for Semiconductor Physics, KU Leuven, B-3001 Leuven, Belgium}
\affiliation{KU Leuven, Department of Electrical Engineering (ESAT), STADIUS Center for Dynamical Systems, Signal Processing and Data Analytics, B-3001 Leuven, Belgium}
\author{Walter Del Pozzo\,\orcidlink{0000-0003-3978-2030}}
\affiliation{Dipartimento di Fisica “E. Fermi”, Università di Pisa, I-56127 Pisa, Italy}
\affiliation{INFN, Sezione di Pisa, I-56127 Pisa, Italy}

\begin{abstract}
    \noindent The Einstein Telescope faces a critical data analysis challenge with correlated noise, often overlooked in current parameter estimation analyses. We address this issue by presenting the statistical formulation of the likelihood that includes correlated noise for the Einstein Telescope or any detector network. By considering varying degrees of correlation, we probe the impact of noise correlations on the parameter estimation analysis of a GW150914-like event. We show that neglecting these correlations may significantly reduce the accuracy of the chirp mass reconstruction. This emphasizes how critical a proper treatment of correlated noise is, as presented in this work, to unlocking the wealth of results promised by the Einstein Telescope.
\end{abstract}

\maketitle

\begin{acronym}
    \acro{GW}[GW]{gravitational-wave}
    \acro{GWs}[GWs]{gravitational-waves}
    \acro{ET}[ET]{Einstein Telescope}
    \acro{PE}[PE]{parameter estimation}
    \acro{PSD}[PSD]{Power Spectral Density}
    \acro{CSD}[CSD]{Cross Power Spectral Density}
    \acro{DFT}[DFT]{Discrete Fourier Transform}
    \acro{SNR}[SNR]{Signal-to-Noise Ratio}
    \acro{BBH}[BBH]{binary black hole}
    \acro{CPSD}[(C)PSD]{(Cross) Power Spectral Density}
    \acro{IMBHs}[IMBHs]{intermediate mass black holes}
    \acro{BNS}[BNS]{binary neutron star}
    \acro{LISA}[LISA]{Laser Interferometer Space Antenna}
\end{acronym}


\section{Introduction}
\noindent The second-generation of \ac{GW} interferometers has detected a large number of coalescing binaries \cite{GW150914, GW170814, GW170817, GW190521, GWTC-3}, leading to groundbreaking scientific results.
The upcoming third-generation of detectors promises to open up new frontiers in the exploration of the Universe, making it possible to address a variety of problems in astrophysics, fundamental physics, and cosmology \cite{ET_science_case, Maggiore_2020_ET_science_case, Branchesi_2023_ET_science_case, CE_horizon_study, LISA_science_case}.
The European proposal for a third-generation ground-based detector is represented by the \ac{ET} \cite{Punturo_2010}.
The \ac{ET}'s proposed design features six-in-one colocated interferometers ---three specialized in the low-frequency observations, three in the high-frequency--- with an opening angle of $\pi/3$, disposed underground in a triangular configuration.\\
\indent Due to the small distance between the input/output test masses of different interferometers \cite{ET_design_report}, non-negligible correlations in the noise are expected among detectors in the \ac{ET}. These correlations arise mainly in the form of magnetic, seismic and Newtonian noise, and could highly limit all kinds of unmodeled searches which rely on cross-correlating data, such as searches for the stochastic \ac{GW} background \cite{Janssens_magnetic_noise, Ball_lightning_strokes, Janssens_newtonian_seismic, Thrane_correlations_SGWB, Christensen_2019_SGWB, Janssens_NN_2024}.
This issue is also well known for the future space-based gravitational wave detector \ac{LISA} \cite{LISA, LISA_redbook}.
Sharing the same triangular geometry as the \ac{ET}, \ac{LISA} will similarly be subject to correlations in instrumental noise which, along with an astrophysical foreground of galactic binaries, will limit its capacity to observe the stochastic background of \ac{GW}s \cite{Cornish_2001, Adams_2010, Boileau_2022, Baghi_2023}.\\
\indent Although the presence of correlated noise has been widely recognized by the \ac{GW} community, it is usually neglected in the context of \ac{PE} analysis, where different detectors forming a network are assumed to be uncorrelated \cite{Veitch_vecchio_Bayesian_PE, Christensen_PE_for_GW}.
Several techniques have been developed to subtract these correlations or mitigate their impact \cite{Coughlin_2016_corr_noise_subtraction, Magnetic_noise_subtraction, Badaracco_2019_NN_subtraction}, improving the detector sensitivity and the subsequent estimation of source parameters \cite{AdLIGO_PE_noise_subtraction, Davis_2019_LIGO_noise_subtraction}.\\

The flagship results for the \ac{ET} necessitate precise \ac{PE} analysis, such as testing General Relativity, probing neutron star equations of state or inferring cosmological parameters.
Similarly, searches for a stochastic background of \ac{GW}s, signals from core-collapse supernovae and rotating neutron stars rely on an accurate identification of resolved signals. Nevertheless, the presence of correlated noise will have a significant impact on achieving this.
Without proper treatment, we may threaten the scientific potential of the \ac{ET}. \\
\indent Differently from the existing literature, we address the issue of including correlated noise directly in the \ac{PE} analysis of \ac{GW} signals.
We consider a statistical derivation of the likelihood, both in its time and frequency domain, for analysing \ac{GW} data with a network of correlated detectors.
We investigate the consequences of neglecting correlated noise in the \ac{ET} by varying the level of correlation for a GW150914-like event.
Importantly, we show a significant reduction in the accuracy of the chirp mass reconstruction.
The main goal of this paper is to emphasize that ignoring correlations in \ac{PE} analysis can undermine the wealth of results expected from the \ac{ET}, stressing the importance of using a likelihood that accounts for correlations.
\vspace{-7pt}

\section{Likelihood Formulation}\label{sec: Likelihood Formulation}
\noindent In the context of \ac{GW} data analysis, one of the pivotal elements is the likelihood $p(\vect{d} | \vect{\theta})$, which represents the probability density function of the observed data $\vect{d}$ given a set of parameters $\vect{\theta}$.
Failing in an accurate construction endangers robust \ac{PE} and hypothesis testing.
The inclusion of correlations in the likelihood formulation has a long history in time series analysis \cite{whittle1953_multivariate, Dunsmuir1979ACL, rao2020, krafty, meier2020104560}, as well as in \ac{GW} data analysis \cite{cutler_flanagan, Adams_2010, Romano_2017, Littenberg_2020, Baghi_2023}.
Nevertheless, a study on the effect of noise correlations on the \ac{PE} of \ac{GW} signals has never been presented.\\
\indent Here, we consider an alternative representation of the standard multivariate Whittle likelihood \cite{whittle1953_multivariate}, which has the advantage of clearly illustrating the nature of the noise in each detector and detector pair, and we refer the interested reader to the discussion in Appendix~\ref{App: Standard and compact multivariate Whittle likelihood representations} for details on the implementation.
For the purpose of the analysis, we treat the \ac{ET} as consisting of three interferometers, ignoring the details of the xylophone configuration \cite{Hild_2011_ET_sensitivity}.
However, the following derivation can be generalized straightforwardly to any network of \ac{GW} interferometers.

\subsection{Likelihood for correlated noise}
\noindent Within a network of \ac{GW} detectors, we indicate with $\vect{n}_\ell$ the noise time series for the $\ell\text{-th}$ interferometer, assuming a sampling frequency of $1/\Delta t$.
In the usual assumption of Gaussian, zero-mean and wide-sense stationary noise, the spatial and temporal correlations between the noise in the $\ell\text{-th}$ and the $m\text{-th}$ interferometers are expressed by the cross-covariance matrix $\vect{\Sigma}_n^{\ell m} \coloneq \mathbb{E} \left[ \vect{n}_\ell \vect{n}_{m}^T\right]$.
In particular, $\vect{\Sigma}_n^{\ell m}$ assumes a Toeplitz form for $\ell \neq m$ and a symmetric Toeplitz form for $\ell=m$.\\
\indent Conventionally, the noise time series of multiple detectors is expressed as a matrix $\vect{N}\in\mathbb{R}^{3\times N}$, where each row $\ell=1,2,3$ corresponds to the time series of the $\ell\text{-th}$ interferometer, and $N$ denotes the number of data points in each time series.
To facilitate the characterization of the spatial and temporal correlation, we introduce the vectorized noise time series defined as follows:
\begin{equation}
    \vect{n} \coloneq \text{vec}(\vect{N}^T) =
                                \begin{bmatrix}
                                    \vect{n}_{1} \\
                                    \vect{n}_{2} \\
                                    \vect{n}_{3}
                                \end{bmatrix},
\end{equation}
where $\vect{n}\in\mathbb{R}^{3N}$.
The spatial and temporal correlations of the noise process $\vect{n}$ are characterized by the $3N\times3N$ network covariance matrix $\vect{\Sigma}_n$, given by
\begin{subequations}\label{eq: covariance matrix}
\begin{align}
  \vect{\Sigma}_n &:= \mathbb{E} \left[ \vect{n}\vect{n}^T \right] \\[1pt]
                & =\begin{bmatrix}[1.1]
                      \mathbb{E}\left[\vect{n}_{1} \vect{n}_{1}^{T}\right] &
                      \mathbb{E}\left[\vect{n}_{1} \vect{n}_{2}^{T}\right] &
                      \mathbb{E}\left[\vect{n}_{1} \vect{n}_{3}^{T}\right] \\
                      \mathbb{E}\left[\vect{n}_{2} \vect{n}_{1}^{T}\right] &
                      \mathbb{E}\left[\vect{n}_{2} \vect{n}_{2}^{T}\right] &
                      \mathbb{E}\left[\vect{n}_{2} \vect{n}_{3}^{T}\right] \\
                      \mathbb{E}\left[\vect{n}_{3} \vect{n}_{1}^{T}\right] &
                      \mathbb{E}\left[\vect{n}_{3} \vect{n}_{2}^{T}\right] &
                      \mathbb{E}\left[\vect{n}_{3} \vect{n}_{3}^{T}\right]
                  \end{bmatrix} \\[1pt]
                &= \begin{bmatrix}[1.3]
                      \vect{\Sigma}_n^{11} & \vect{\Sigma}_n^{12} & \vect{\Sigma}_n^{13} \\
                      \vect{\Sigma}_n^{21} & \vect{\Sigma}_n^{22} & \vect{\Sigma}_n^{23} \\
                      \vect{\Sigma}_n^{31} & \vect{\Sigma}_n^{32} & \vect{\Sigma}_n^{33}
                    \end{bmatrix}.
\end{align}
\end{subequations}
\vfill\eject
\noindent In particular, the diagonal blocks characterize the temporal correlation of the noise process in each detector, and the off-diagonal blocks characterize the spatial correlation of the noise processes between different detectors.
For the off-diagonal blocks $\vect{\Sigma}_n^{\ell m} = (\vect{\Sigma}_n^{m\ell})^T$.\\

Consider the hypothesis that the data $\vect{d}$ recorded by the \ac{ET} contain a \ac{GW} signal that depends on a set of source parameters $\vect{\theta}$.
We model the output as
\begin{equation}
    \vect{d} = \vect{s}(\vect{\theta}) + \vect{n}
\end{equation}
where $\vect{s}(\vect{\theta})$ and $\vect{n}$ are the \ac{GW} signal and the noise component, respectively.
For such a hypothesis, the likelihood of observing the data $\vect{d}$ follows the distribution of the noise, that is
\begin{equation}\label{eq: likelihood network time}
    p(\vect{d}|\vect{\theta}) = \dfrac{1}{|2\pi\vect{\Sigma}_n|^{1/2}} \exp{\bigg[ \hspace{-3pt} -\frac{1}{2} (\vect{d}}-\vect{s})^T \vect{\Sigma}_n^{-1} (\vect{d}-\vect{s}) \bigg],
\end{equation}
which corresponds to the maximum entropy distribution of the zero-mean noise processes constrained by the network covariance matrix $\vect{\Sigma}_n$ \cite{Gregory_2005}.\\

Although the time-domain formulation is essential for understanding the data as it is recorded, given the reduced complexity it is useful to present the frequency-domain version of Eq.~\eqref{eq: likelihood network time}.
Let $\vect{S}_n$ be the spectral matrix defined as
\begin{equation}\label{eq: PSD network matrix}
  \vect{S}_n \coloneq \begin{bmatrix}[1.3]
                      \vect{S}_n^{11} & \vect{S}_n^{12} & \vect{S}_n^{13} \\
                      \vect{S}_n^{21} & \vect{S}_n^{22} & \vect{S}_n^{23} \\
                      \vect{S}_n^{31} & \vect{S}_n^{32} & \vect{S}_n^{33}
                    \end{bmatrix},
\end{equation}
where
\begin{subequations}
\begin{align}
    &S_{n}^{\ell m}(f_k) = 2\Delta f \hspace{3pt} \delta_{jk} \hspace{3pt} \mathbb{E}\left[\tilde{n}_{\ell}(f_j)\tilde{n}_{m}^{*}(f_k)\right]\\[3pt]
    &\vect{S}_n^{\ell m} = \text{diag}\left[S_n^{\ell m}(f_{k_\text{low}}),\dots,S_n^{\ell m}(f_{k_\text{high}})\right], \label{eq:PSD matrix}
\end{align}
\end{subequations}
with the tilde indicating the Fourier transform and ~~$\Delta f = 1/N\Delta t$ the frequency resolution.
In particular, $S_n^{\ell\ell}(f_k)$ is the $k\text{-th}$ component of the one-sided \ac{PSD} for the $\ell\text{-th}$ interferometer, and $S_n^{\ell m}(f_k)$ the $k\text{-th}$ component of the one-sided \ac{CSD} between the $\ell\text{-th}$ and the $m\text{-th}$ interferometers.
Note that for the off-diagonal blocks $\vect{S}_n^{\ell m} = (\vect{S}_n^{m \ell})^*$.\\
\indent It can be shown that each (cross-)covariance Toeplitz matrix $\vect{\Sigma}_n^{\ell m}$ is asymptotically equivalent to a circulant matrix, meaning that its eigenvalues are the \ac{DFT} of the first column \cite{Toeplitz_and_Circulant_Matrices_Grey, Toeplitz_asymptotics}.
In other words,  each block of the covariance matrix $\vect{\Sigma}_n$ in Eq.~\eqref{eq: covariance matrix} can be independently diagonalized by the \ac{DFT} basis, resulting in the corresponding (Cross) \ac{PSD} matrix, Eq.~\eqref{eq:PSD matrix}.
This yields the following frequency-domain representation of the likelihood \cite{whittle1953_multivariate}:
\begin{widetext}
\begin{equation}\label{eq: likelihood network frequency 1}
    p(\vect{d}|\vect{\theta}) \approx
        \dfrac{1}{|\pi\vect{S}_n/2\Delta f|}
        \exp \left[
            -2\Delta f
            \left( \tilde{\vect{d}} - \tilde{\vect{s}}(\vect{\theta})\right)^\dagger
            \hspace{3pt} \vect{S}_n^{-1} \hspace{3pt}
            \left( \tilde{\vect{d}} - \tilde{\vect{s}}(\vect{\theta})\right)
            \right],
\end{equation}
\end{widetext}
where the dagger $(\dagger)$ represents the conjugate transpose, and the tilde on $\tilde{\vect{d}}$ denotes that the \ac{DFT} is applied on each individual row of the matrix $\vect{D}\in\mathbb{R}^{3\times N}$ (the same applies for $\tilde{\vect{s}}$).
Since the entries of $\vect{S}_n$ are diagonal matrices, the inverse of $\vect{S}_n$ also follows the same structure (see Appendix~\ref{App: Standard and compact multivariate Whittle likelihood representations} for a proof). A discussion on the spectral matrix inversion is also provided in Appendix~\ref{App: Standard and compact multivariate Whittle likelihood representations}.
\\
\indent One can rewrite Eq.~\eqref{eq: likelihood network frequency 1} in the more compact form
\begin{equation}\label{eq: likelihood network frequency 2}
    p(\vect{d}|\vect{\theta}) \approx
        \dfrac{1}{|\pi\vect{S}_n/2\Delta f|} \hspace{3pt}
        e^{- \frac{1}{2} \left(\vect{d} - \vect{s}(\vect{\theta})\hspace{1pt} | \hspace{1pt} \vect{d} - \vect{s}(\vect{\theta})\right)}
\end{equation}
where $\big(\vect{x} \hspace{1pt} | \hspace{1pt} \vect{x} \big)$ is the noise-weighted inner product of the network time series $\vect{x}$ with itself, defined as
\begin{equation}\label{eq: inner product}
    \big(\vect{x} \hspace{1pt} | \hspace{1pt} \vect{x} \big) \coloneq
        4\Delta f \hspace{2pt} \Re
            \left[ \sum_{\ell,m=1}^{3}\sum_{k} \tilde{x}_{\ell}^{*}(f_k)
            \hspace{3pt} (\vect{S}_n^{-1})^{\ell m}(f_k) \hspace{3pt}
            \tilde{x}_{m}(f_k)
            \right].
\end{equation}\\
In particular, $(\vect{S}_n^{-1})^{\ell m}(f_k)$ denotes the $k\text{-th}$ diagonal entry of the block $\ell m$ of the inverse of $\vect{S}_n$.
The index $k$ varies from the DC frequency at $k = 0$ to the Nyquist frequency at $k = N/2-1$, both extremes excluded.

\subsection{Likelihood for uncorrelated noise}

\noindent In the case of uncorrelated detectors, the off-diagonal blocks of the network covariance matrix $\vect{\Sigma}_n$ and of the spectral matrix $\vect{S}_n$ vanish, that is
\begin{subequations}
\begin{align}
    \vect{\Sigma}_n^{\text{uncorr}} &\coloneq \begin{bmatrix}
                                              \vect{\Sigma}_n^{11} &     \vect{0}       &     \vect{0}       \\
                                                  \vect{0}       & \vect{\Sigma}_n^{22} &     \vect{0}       \\
                                                  \vect{0}       &     \vect{0}       & \vect{\Sigma}_n^{33}
                                            \end{bmatrix},\\
    \nonumber \\[-10pt]
    \vect{S}_n^{\text{uncorr}} &\coloneq \begin{bmatrix}
                                          \vect{S}_n^{11} &   \vect{0}    &   \vect{0}     \\
                                            \vect{0}    & \vect{S}_n^{22} &   \vect{0}     \\
                                            \vect{0}    &   \vect{0}    & \vect{S}_n^{33}
                                        \end{bmatrix}. \label{eq: PSD network matrix uncorr}
\end{align}
\end{subequations}
For this specific case, the time-domain representation of the likelihood in Eq.~\eqref{eq: likelihood network time} is simply given by the product of the single detector likelihoods:
\begin{equation}
    p(\vect{d}|\vect{\theta}) \propto
        \prod_{\ell=1}^3
            \exp{\left[ -\frac{1}{2}
                \left(\vect{d}_\ell-\vect{s}_\ell\right)^T
                \left(\vect{\Sigma}_n^{\ell\ell}\right)^{-1}
                \left(\vect{d}_\ell-\vect{s}_\ell\right)
            \right]}.
\end{equation}
Similarly, the frequency-domain representation in Eq.~\eqref{eq: likelihood network frequency 1} reduces to the product of the univariate Whittle likelihoods for each detector $\ell$ \cite{whittle1953_estimation}, which is the form currently used in \ac{PE} analysis with second-generation detectors:
\begin{equation}\label{eq: likelihood network uncorrelated}
    p(\vect{d}|\vect{\theta}) \propto
        \prod_{\ell=1}^3
            \exp{\left[ -2\Delta f
                \left(\tilde{\vect{d}_\ell}-\tilde{\vect{s}}_\ell\right)^\dagger
                \left(\vect{S}_n^{\ell\ell}\right)^{-1}
                \left(\tilde{\vect{d}_\ell}-\tilde{\vect{s}}_\ell\right)
            \right]}.
\end{equation}

\section{Parameter Estimation in the presence of correlated noise}\label{Sec: Parameter Estimation in the presence of correlated noise}
\noindent To demonstrate the impact of ignoring noise correlations, we perform several \ac{PE} analyses with the \ac{ET} using the \textit{correlated} likelihood, Eq.~\eqref{eq: likelihood network frequency 1}, and the \textit{uncorrelated} likelihood, Eq.~\eqref{eq: likelihood network uncorrelated}, investigating how a \ac{GW} signal is reconstructed by the two models.
In particular, we consider a single \ac{BBH} signal, and we repeat the \ac{PE} analysis varying the level of correlation between the interferometers.
The simulations are performed using the software package \texttt{JIM} \cite{JIM_2023}.

\subsection{Simulations setup}
\noindent In a realistic scenario for the \ac{ET}, the spectral matrix $\vect{S}_n$ is composed of the estimated \ac{PSD}s for each interferometer and of the estimated \ac{CSD}s for each interferometer pair.
Estimating the \ac{PSD}s involves autocorrelating the strain from each individual detector; similarly, to obtain an estimate of the \ac{CSD}s all that is needed, in principle, is to cross-correlate the strain of data measured by each detector.\\

In our analyses, we assume the same \ac{PSD} for every detector, equivalent to the design sensitivity of the \ac{ET} xylophone configuration \cite{Danilishin_Zhang}, $S_n^{\text{ET}}$, i.e.
\begin{equation}\label{eq:PSD_ET}
    S_n^{\ell\ell}(f) = S_n^{\text{ET}}(f), \hspace{20pt} \ell=1,2,3.
\end{equation}
The \ac{CSD} can be expressed in terms of the correlation coefficients as follows:
\begin{subequations}\label{eq:CSD_ET}
\begin{align}
    S_n^{\ell m}(f) &= \alpha_{\ell m}(f) \hspace{2pt} \sqrt{S_n^{\ell\ell}(f) \hspace{3pt} S_n^{mm}(f)}\\
                 &= \alpha_{\ell m}(f) \hspace{2pt} S_n^{\text{ET}}(f)
\end{align}
\end{subequations}
where $\alpha_{\ell m}(f)\in\mathbb{C}$ describes the correlation between the detectors $\ell$ and $m$.
Since noise correlations are likely to arise in the form of seismic, Newtonian and magnetic noise, which are low frequency noise components \cite{Janssens_magnetic_noise, Ball_lightning_strokes, Janssens_newtonian_seismic, Janssens_NN_2024}, we assume in our simulations that the interferometers are correlated only for frequencies below $10$ Hz.
To investigate the impact of noise correlations, in the absence of a faithful model, we vary $\alpha_{\ell m}(f)$ in the range $\left[-0.5, 0.9\right]$ in steps of $0.1$, assuming the same correlation coefficient for each interferometer pair, that is
\begin{equation}\label{eq:alpha}
\alpha_{\ell m}(f) = \alpha(f) \in \left[-0.5, 0.9\right] \hspace{20pt} \text{for } f \leq 10\text{ Hz}
\end{equation}
and $\alpha_{\ell m}(f)=0$ otherwise.
In particular, the lower bound of $\alpha(f)=-0.5$ is fixed by the assumption that the \ac{PSD}s and the correlation coefficients are equivalent for the three detectors and detector pairs (see Appendix~\ref{App: Lower limit for the correlation coefficient} for a proof).\\

With the expressions in Eq.~\eqref{eq:PSD_ET} -~\eqref{eq:alpha} we construct, for each level of correlation $\alpha = -0.5, -0.4, \dots, 0.9$, the spectral matrix $\vect{S}_n$, as in Eq.~\eqref{eq: PSD network matrix}, and the uncorrelated version $\vect{S}_n^{\text{uncorr}}$, defined in Eq.~\eqref{eq: PSD network matrix uncorr}:

\begin{subequations}
\begin{align}
    \vect{S}_{n} &=
    \begin{bmatrix}[1.3]
        \vect{S}_{n}^\text{ET}       & \alpha\vect{S}_{n}^\text{ET} & \alpha\vect{S}_{n}^\text{ET} \\
        \alpha\vect{S}_{n}^\text{ET} & \vect{S}_{n}^\text{ET}       & \alpha\vect{S}_{n}^\text{ET} \\
        \alpha\vect{S}_{n}^\text{ET} & \alpha\vect{S}_{n}^\text{ET} & \vect{S}_{n}^\text{ET}
    \end{bmatrix}, \label{eq: spectral matrix corr} \\
    \nonumber \\[-10pt]
    \vect{S}_{n}^\text{uncorr} &=
    \begin{bmatrix}
        \vect{S}_{n}^\text{ET} & \vect{0}               & \vect{0} \\
        \vect{0}               & \vect{S}_{n}^\text{ET} & \vect{0} \\
        \vect{0}               & \vect{0}               & \vect{S}_{n}^\text{ET}
    \end{bmatrix}. \label{eq: spectral matrix uncorr}
\end{align}
\end{subequations}

\noindent For every simulation, we produce the mock data for the \ac{ET} generating the correlated noise from a zero-mean multivariate normal distribution, with $\vect{S}_n/2$ as the covariance matrix.
The real and imaginary parts are generated independently and summed to obtain the network frequency series $\tilde{\vect{n}}$.
Precisely, we select a sampling frequency of $1/\Delta t = 2048\text{ Hz}$, and we discard all the frequencies below $5\text{ Hz}$ and above the Nyquist frequency $1/2\Delta t$. Finally, we generate a synthetic \ac{BBH} signal with the \texttt{IMRPhenomD} frequency-domain model \cite{IMRPhenomD}, and we inject it into the noise frequency series.
In all \ac{PE} analyses, we assume the spectral matrix of noise to be known through Eqs.~\eqref{eq: spectral matrix corr} and \eqref{eq: spectral matrix uncorr}.

\subsection{Network \ac{SNR} with correlated noise}
\noindent In the context of matched filtering, the optimal value of the \ac{SNR} for a single detector analysis is defined as
\begin{equation}
    \rho^\ell =  \sqrt{4\Delta f \hspace{4pt} \Re \left( \sum_{k} \dfrac{|\tilde{s}_{\ell}(f_k)|^2}{S_n^{\ell\ell}(f_k)} \right)}
\end{equation}
where $\tilde{s}_{\ell}(f_k)$ is the $k\text{-th}$ frequency component of the \ac{GW} signal detected in the $\ell\text{-th}$ interferometer \cite{Veitch_vecchio_Bayesian_PE, Christensen_PE_for_GW}.
From Eq.~\eqref{eq: inner product}, we generalize the expression to the case where correlated noise is present as follows:
\begin{equation}\label{eq: network SNR}
    \rho^\text{corr} \coloneq  \left(\vect{s} \hspace{1pt} | \hspace{1pt} \vect{s} \right)^{1/2},
\end{equation}
where $\vect{s}(\vect{\theta})$ is the network time series of the detected \ac{GW} signal \cite{cutler_flanagan}.
In particular, we expect a better efficacy of spectral component weighting when accounting for correlated noise, resulting in a more accurate calculation of the \ac{SNR}.
This behaviour is due to the nature of correlated noise itself, which enhances the collective information about noise processes akin to a network of witness sensors measuring the same physical phenomenon.
The corresponding values of $\rho^\text{corr}$ for the event considered and the correlation levels used in our simulations are presented in Sec.~\ref{sec: Results}, while the impact of the correlation coefficient’ sign on the SNR is discussed in Sec.~\ref{Sec: Effects of correlated noise on the signal space}.
\\
\indent Note once again that, in the case of uncorrelated detectors, Eq.~\eqref{eq: network SNR} reduces to the standard expression used in current \ac{GW} analyses, given by the quadrature sum of the \ac{SNR} for each detector,
\begin{equation}\label{eq: network SNR uncorr}
    \rho^\text{uncorr} = \sqrt{\sum_{\ell=1}^3 (\rho^\ell)^2 }
\end{equation}
For the event considered in our simulations, we have $\rho^\text{uncorr}=20.0$, which remains fixed for every $\alpha$ as it only depends on the diagonal blocks of the spectral matrix through Eq.~\eqref{eq: spectral matrix uncorr}.

\subsection{Results}\label{sec: Results}
\begin{figure*}[t!]
    \includegraphics[width=\linewidth]{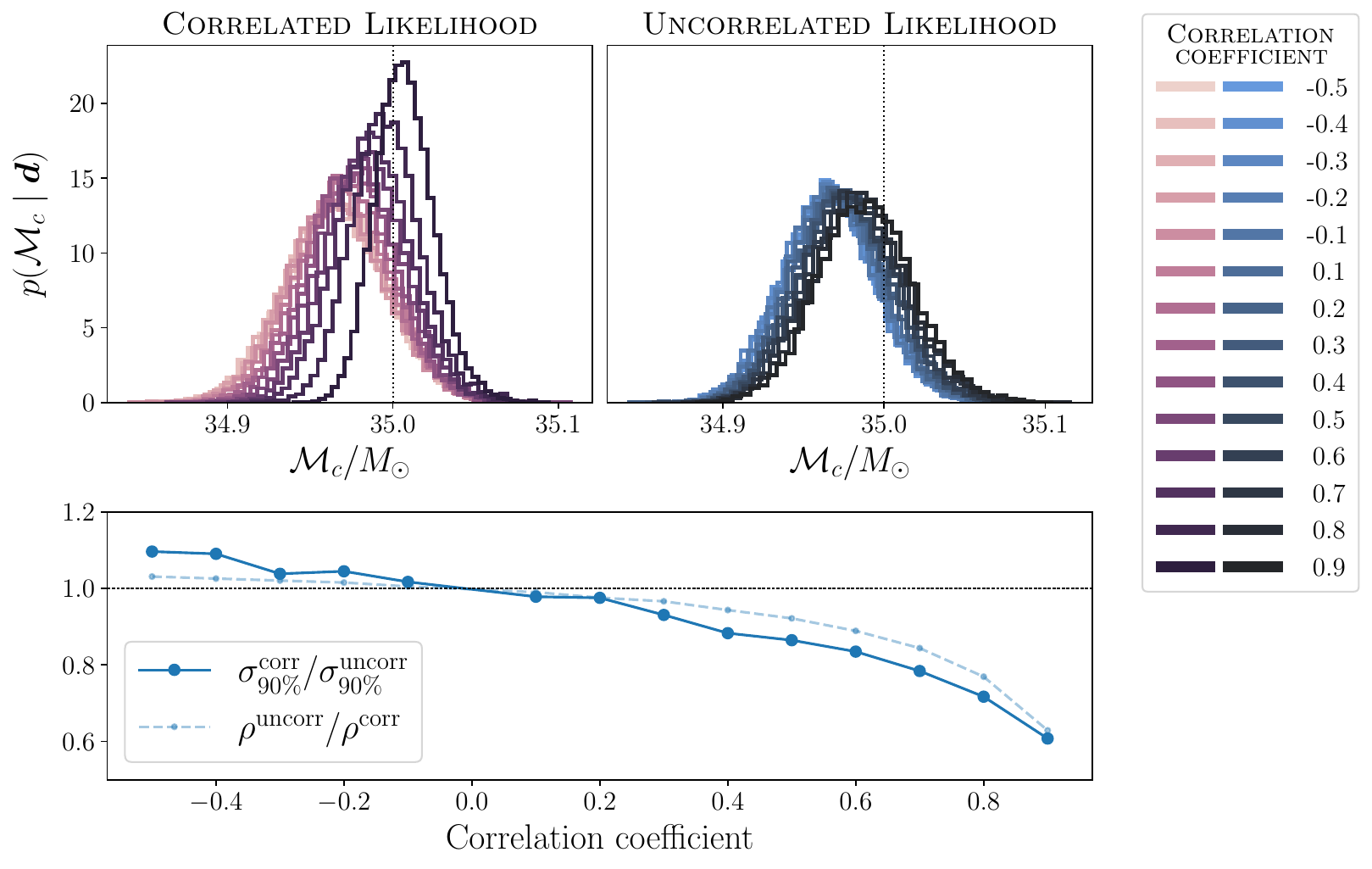}
    \caption{(\textbf{top}) Chirp mass posterior distribution reconstructed with the correlated and uncorrelated likelihoods for different values of the correlation coefficient $\alpha$. The dashed lines represent the injected value. The reduction in the spread of the distributions due to the presence of correlated noise is not captured in the analyses performed with the uncorrelated likelihood. (\textbf{bottom}) The solid line represents the ratio of the $90\%$ credible interval for the chirp mass posterior distributions reconstructed with the correlated likelihood over the uncorrelated likelihood, as a function of the correlation coefficient $\alpha$. For comparison, the ratio between the uncorrelated and correlated \ac{SNR} as a function of the correlation coefficient $\alpha$ is reported as a dashed line. The trend of $\rho^\text{corr}$ can reasonably explain the narrowing of the chirp mass posterior distributions with increasing $\alpha$.}
    \label{fig: chirp_mass_comparison}
    \vspace{20pt}
\end{figure*}

\begin{figure*}[t]
\begin{center}
    \centering
    \includegraphics[width=1.0\linewidth]{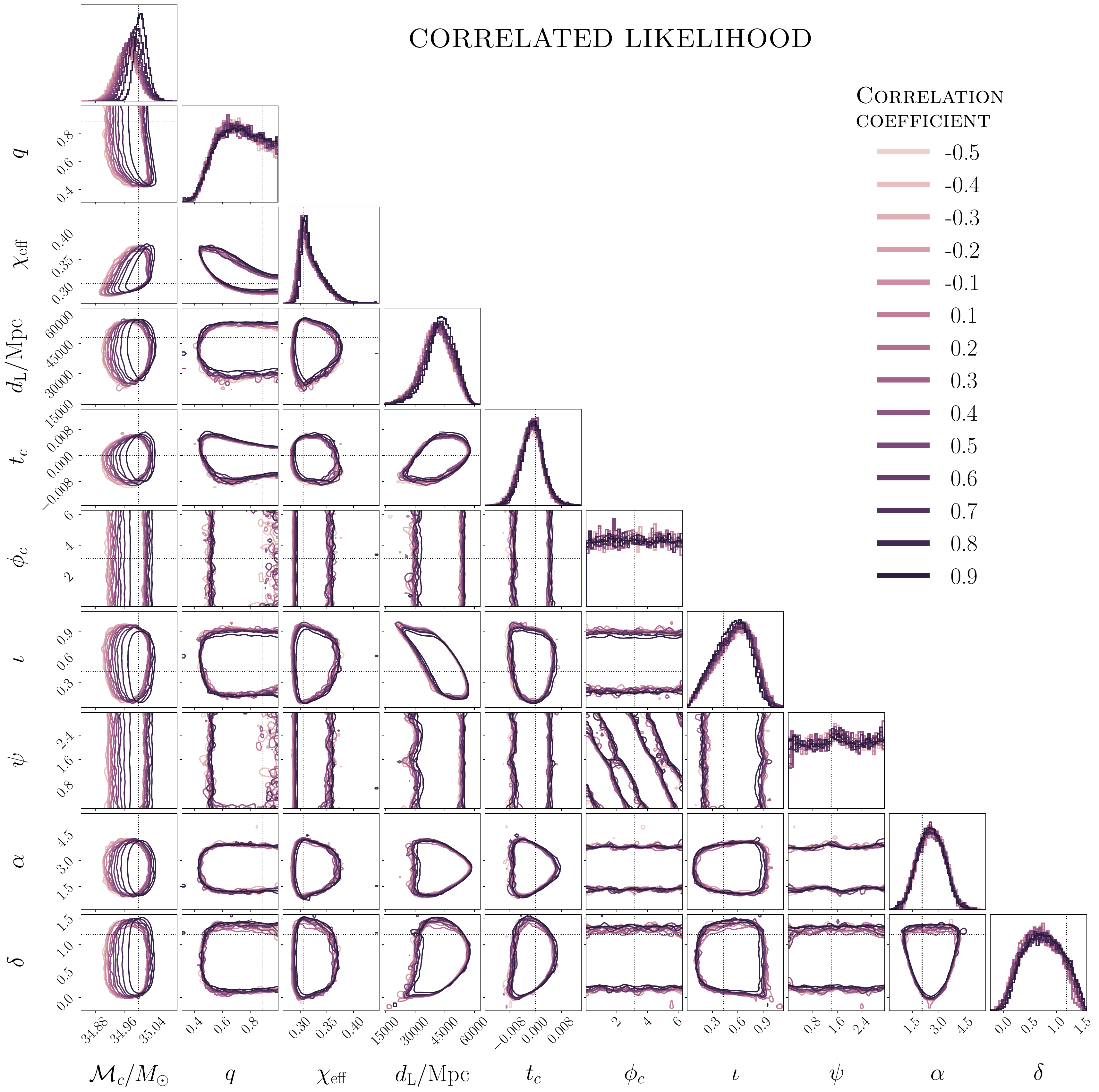}
     \vspace{0.2cm}
    \caption{Posterior distributions for all parameters from the PE analyses performed using the correlated likelihood, with varying levels of the correlation coefficient $\alpha$. The contours in the 2D plots enclose $90\%$ of the probability mass, and the dashed lines represent the injected parameter values.}
    \label{fig: correlated corner plot}
\end{center}
\end{figure*}

\begin{figure*}[t]
\begin{center}
    \centering
    \includegraphics[width=1.0\linewidth]{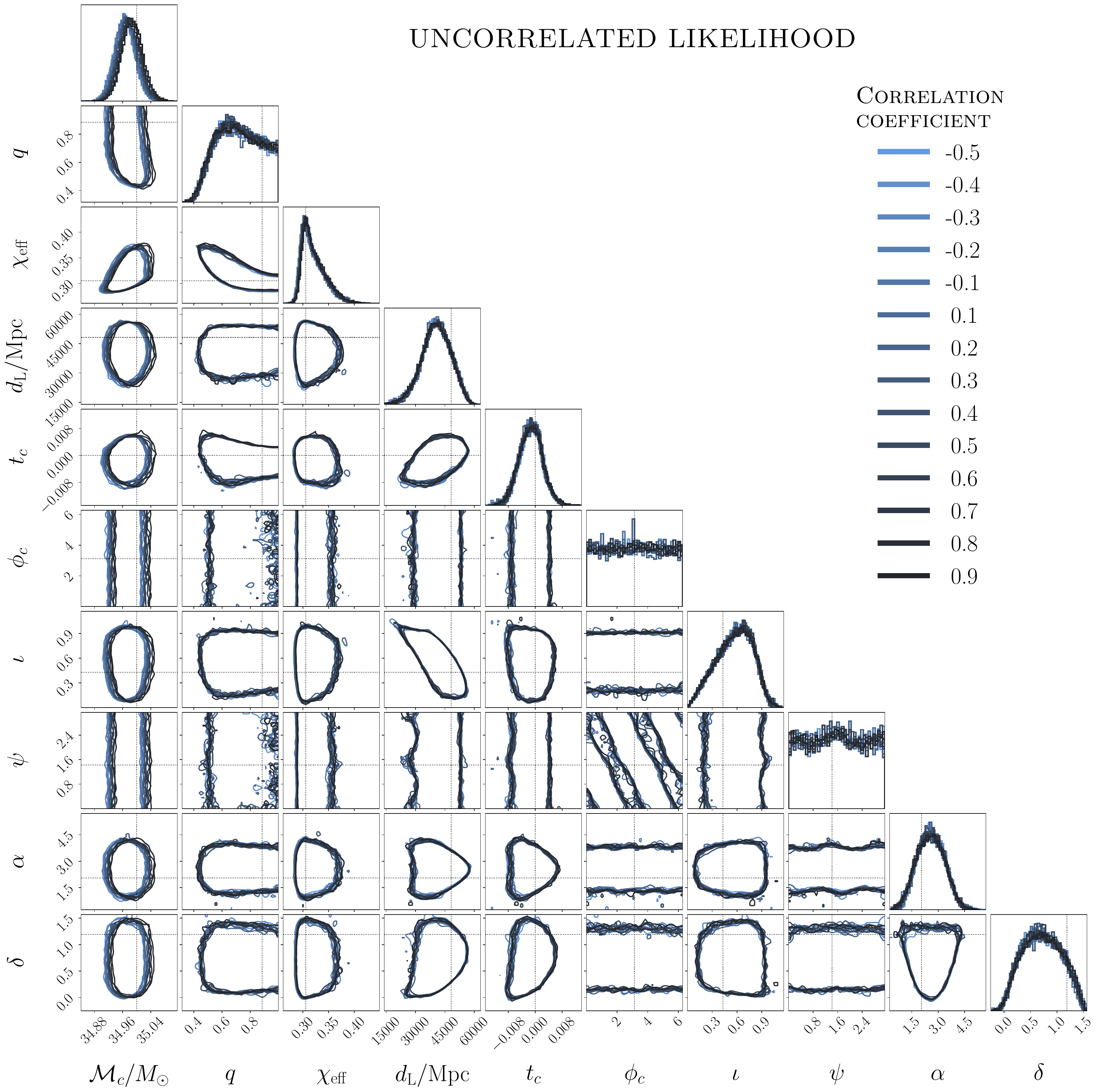}
     \vspace{0.2cm}
    \caption{Posterior distributions for all parameters from the PE analyses performed using the uncorrelated likelihood, with varying levels of the correlation coefficient $\alpha$. The contours in the 2D plots enclose $90\%$ of the probability mass, and the dashed lines represent the injected parameter values.}
    \label{fig: uncorrelated corner plot}
\end{center}
\end{figure*}

\noindent The simulations indicate a significant difference in the chirp mass reconstruction when using the correlated or the uncorrelated likelihood.\\
\indent In Fig.~\ref{fig: chirp_mass_comparison} we compare the spread of the chirp mass posterior distribution for the two likelihoods as the correlation coefficient $\alpha$ increases.
The top-left panel shows that correlated noise clearly impacts the spread of the distributions, which decreases as $\alpha$ increases.
This effect is not captured when using the uncorrelated likelihood, as shown in the top-right panel, where the posterior distribution remains almost unchanged despite varying the level of correlation.\\
\indent To quantify this effect, we report the ratio of the spreads for the correlated and the uncorrelated cases in the bottom panel.
The spread ratio moves away from $1$ as $|\alpha|$ increases, reaching a remarkable deviation of $60\%$ for high levels of correlations $(\alpha = 0.9)$.
Specifically, the $\mathcal{M}_c$ posterior distributions reconstructed with the correlated likelihood are broader compared to the uncorrelated case when $\alpha<0$ and narrower when $\alpha>0$.
For comparison, the ratio between the uncorrelated injected \ac{SNR}, Eq.~\eqref{eq: network SNR uncorr}, and the correlated injected \ac{SNR}, Eq.~\eqref{eq: network SNR}, is shown as a dashed line.
The trend of $\rho^\text{corr}$, smaller than $\rho^\text{uncorr}$ for negative correlations and larger for positive correlations, can reasonably explain the narrowing of the chirp mass posterior distributions with increasing $\alpha$.
Given the close relationship between the effects of correlations and the changes in the \ac{SNR}, we can expect a similar behaviour for other \ac{BBH} events or \ac{GW} sources.
Nevertheless, for the event considered, the impact of correlated noise on the reconstruction of the other parameters is not as pronounced, as shown in Figs.~\ref{fig: correlated corner plot} and \ref{fig: uncorrelated corner plot}.\\

The analysis highlights the potential risks of using the uncorrelated likelihood in Eq.~\eqref{eq: likelihood network uncorrelated} for the \ac{ET}, leading to overconfident parameter reconstruction with negative correlations, and underconfident reconstruction with positive correlations.
This inaccuracy, which stems from neglecting correlated noise, may introduce biases in the results and reduce the ability to constrain the parameters effectively.

\subsection{Effects of correlated noise on the signal space}\label{Sec: Effects of correlated noise on the signal space}
\noindent To better understand the observed effects of correlated noise on the \ac{PE} analyses, it is helpful to consider a different representation.
One key advantage of the triangular configuration is the possibility to decompose the observational space of the \ac{ET} into the so-called \textit{principal coordinate system}, consisting of a one-dimensional null space and a two-dimensional signal space. As presented in \cite{Wong_2022}, by applying the rotation matrix $\vect{U}$,
\begin{equation}
    \vect{U}=
    \begin{bmatrix}
        -\sqrt{6}/6 & -\sqrt{2}/2 & \sqrt{3}/3 \\
         \sqrt{6}/3 &  0          & \sqrt{3}/3 \\
        -\sqrt{6}/6 &  \sqrt{2}/2 & \sqrt{3}/3
    \end{bmatrix},
\end{equation}
\noindent a vector $\vect{x}\in\mathbb{R}^{3N}$ can be transformed into the principal coordinate system as
\begin{equation}
    \vect{x}^p = \vect{U}^T \vect{x} = \begin{bmatrix}[1.3]
                                        \vect{x}_1^{p,T}\\
                                        \vect{x}_2^{p,T}\\
                                        \vect{n}^{p,T}
                                        \end{bmatrix}
\end{equation}
where $\vect{x}_1^{p}$, $\vect{x}_2^{p}$ represent the time series of the equivalent detectors in the two-dimensional signal space, and $\vect{n}^{p}$ represents the time series of the null stream in the one-dimensional null space, which contains no \ac{GW} signals \cite{Goncharov_2022}.
This principal coordinate system is equivalent to the $A,E$ and $T$ channels commonly used in the \ac{LISA} community~\cite{Adams_2010, Littenberg_2020}.\\
\indent Assume now that for a certain frequency $f_k$, the spectral matrix has the form
\begin{equation}\label{eq: spectral matrix f_k}
    \vect{S}_n(f_k) =
    \begin{bmatrix}
        1 & \alpha_k & \alpha_k \\
        \alpha_k & 1 & \alpha_k \\
        \alpha_k & \alpha_k & 1
    \end{bmatrix}.
\end{equation}
Transforming this matrix into the principal coordinate system yields:
\begin{subequations}\label{eq: spectral matrix signal-null space}
\begin{align}
    \vect{S}^p_n(f_k) & = \vect{U}^T \vect{S}_n(f_k) \vect{U} \\
                                   & =   \begin{bmatrix}
                                            1-\alpha_k & 0 & 0 \\
                                            0 & 1-\alpha_k & 0 \\
                                            0 & 0 & 1+2\alpha_k
                                        \end{bmatrix}.
\end{align}
\end{subequations}
As we can see, in the assumption of Gaussian noise, the spread of the likelihood $p(\vect{d}^p(f_k)|\vect{\theta})$ in the two-dimensional signal space depends on the sign of the correlation coefficient $\alpha_k$, and it is proportional to $\sqrt{|\text{diag}(1-\alpha_k,1-\alpha_k)|}$.
Specifically, compared to the uncorrelated case ($\alpha_k=0$), the spread of the likelihood increases when $\alpha_k<0$ and decreases when $\alpha_k>0$.\\

This example provides an explanation for the behaviour of $\rho^\text{corr}$ in our analysis, which decreases for negative correlations and increases for positive correlations, along with its subsequent impact on the spread of the chirp mass posterior distribution.\\
\indent It is however important to underline that from Eq.~\eqref{eq: spectral matrix signal-null space} one cannot immediately conclude that the \ac{PE} improves for positive correlations and worsens for negative, as the whole analysis strongly relies on an accurate estimation of the \ac{PSD}s and \ac{CSD}s, which were assumed to be known in our study.
In this sense, the impact of the sign of $\alpha$ on the null space should also be considered, as it could indicate an enhanced (or diminished) capacity to estimate the noise through the null stream \cite{Goncharov_2022, janssens2023formalism}.
Moreover, from Fischer's inequality \coloredcite[blue]{hornMatrixAnalysis1987}, the overall spread of the likelihood, which is proportional to $\sqrt{|\vect{S}_n|}$, is reduced with respect to the zero-correlation case as long as there are any nonzero components of correlation, regardless of the specific structure of the spectral matrix or the sign of $\alpha$.
We emphasize that a comprehensive investigation is necessary to fully assess the impact of correlated noise on \ac{PE} analyses for the \ac{ET}.

\section{Conclusions}\label{Sec: Conlcusion}
\noindent In the era of the \ac{ET}, handling correlated noise is key to unlocking its full scientific potential.
We have presented the time and frequency-domain likelihoods for a network of \ac{GW} interferometers when correlated noise is present.
Analysing a GW150914-like event, we have shown that the accuracy of the \ac{PE} is significantly reduced when ignoring such correlations.
Specifically, this results in a broader (narrower) posterior reconstruction for the chirp mass in the case of positive (negative) correlations.\\
\indent This work shows the crucial role of accounting for correlations in \ac{PE} analysis, enabling precise studies such as tests of General Relativity, neutron star equations of state reconstruction, and cosmological parameter inference.
Neglecting correlations may also lead to imprecise identification of resolvable signals, critical for searches of stochastic \ac{GW}s background, core-collapse supernovae, and rotating neutron stars.
In that sense, the issue might also be relevant to \ac{LISA}, where \ac{PE} analysis is conducted on an unknown number of signals from different sources simultaneously, in what is known as the “global fit” \cite{Littenberg_2020, Littenberg_2023}.\\
\indent Our analysis envisages the necessity for a thorough investigation of the impact of correlated noise on \ac{PE}, incorporating a refined model for noise correlations and exploring different \ac{SNR} values.
Special attention must also be given to different \ac{GW} sources, particularly those more sensitive to the low frequency region where correlations are expected \cite{Janssens_magnetic_noise, Ball_lightning_strokes, Janssens_newtonian_seismic, Janssens_NN_2024}, such as mergers of intermediate-mass black holes (whose chirp frequency is $\lesssim 10$ Hz) and binary neutron star coalescences (which will remain in the \ac{ET} low frequency band for hours) \cite{ET_science_case, ET_design_report}.\\
\indent Future work will focus on assessing the full impact of the correlated versus the uncorrelated likelihood on the \ac{ET} science cases. We advocate abandoning the assumption of uncorrelated detectors which does not fully exploit the information content in the data (such as their correlations) by using the correlated likelihood in Eq.~\eqref{eq: likelihood network frequency 1}.

\begin{acknowledgments}
\noindent The authors would like to thank Justin Janquart and Jan Harms for insightful discussions. This work was partially supported by the Research Foundation - Flanders (Grant No. I002123N). P.T.H.P is supported by the research program of the Netherlands Organization for Scientific Research (NWO). M.W. is supported by the Research Foundation - Flanders (FWO) through Grant No. 11POK24N.
\end{acknowledgments}

\bibliography{bib}

\appendix
\section{Standard and compact representations for the multivariate Whittle likelihood}\label{App: Standard and compact multivariate Whittle likelihood representations}
\noindent We discuss here the equivalence between the standard Whittle likelihood representation \cite{whittle1953_multivariate} and the compact representation considered in this work.\\

In a single detector analysis, the frequency-domain approximation of the likelihood (i.e., the univariate Whittle likelihood \cite{whittle1953_estimation}) is generally used to reduce the complexity, specifically to avoid the inversion of the covariance matrix.
Since the noise is assumed to be stationary, the \ac{PSD} matrix is diagonal, making the inversion straightforward.
The extension from the univariate to the multivariate case takes essentially the same form \cite{Calder1997}.\\
\indent Following the compact representation considered in this work, the spectral matrix $\vect{S}_n$ takes the subsequent form for a three detectors analysis:
\begin{equation}
  \vect{S}_n \coloneqq
  \begin{bmatrix}[1.3]
    \vect{S}_n^{11} &
    \vect{S}_n^{12} &
    \vect{S}_n^{13} \\
    \vect{S}_n^{21} &
    \vect{S}_n^{22} &
    \vect{S}_n^{23} \\
    \vect{S}_n^{31} &
    \vect{S}_n^{32} &
    \vect{S}_n^{33}
  \end{bmatrix},
\end{equation}
where each block $\vect{S}^{\ell m}_n$ is a diagonal matrix.
Compared to the standard Whittle likelihood, this block representation has the advantage of clearly illustrating the nature of the noise in each detector and detector pair, making the expression for the correlated likelihood compact and straightforward.
While the inversion of a $3N\times 3N$ matrix nominally involves $\mathcal{O}(N^{3})$ computations~\cite{farebrother2018linear},  the inverse of $\vect{S}_n$ can be computed analytically with the following form:
\begin{widetext}
  \begin{equation}\label{eq: App spectral matrix inversion analytical}
    \vect{S}_{n}^{-1} =
    \begingroup
    \setlength\arraycolsep{8pt}
    \begin{bmatrix}[1.5]
        (\vect{S}_{n}^{-1})^{11} &
        -(\vect{S}_{n}^{-1})^{11}\vect{\rho}_{1} &
        -(\vect{S}_{n}^{-1})^{11}\vect{\rho}_{2} \\
        -\vect{\rho}_{1}^{\dagger}(\vect{S}_{n}^{-1})^{11} &
        (\vect{S}_{n}^{22/33})^{-1} + \vect{\rho}_{1}^{\dagger} (\vect{S}_{n}^{-1})^{11}\vect{\rho}_{1} &
        -(\vect{S}_{n}^{22/33})^{-1}\vect{\sigma} + \vect{\rho}_{1}^{\dagger}(\vect{S}_{n}^{-1})^{11}\vect{\rho}_{2} \\
        -\vect{\rho}_{2}^{\dagger}(\vect{S}_{n}^{-1})^{11} &
        -\vect{\sigma}^{\dagger}(\vect{S}_{n}^{22/33})^{-1} + \vect{\rho}_{2}^{\dagger}(\vect{S}_{n}^{-1})^{11}\vect{\rho}_{1} &
        (\vect{S}_{n}^{33})^{-1} + \vect{\sigma}^{\dagger} (\vect{S}_{n}^{22/33})^{-1}\vect{\sigma} + \vect{\rho}_{2}^{\dagger} (\vect{S}_{n}^{-1})^{11} \vect{\rho}_{2}
    \end{bmatrix}
  \endgroup
  \end{equation}
where
\begin{align}
  \vect{S}_{n}^{22/33} &= \vect{S}_{n}^{22} - \vect{S}_{n}^{23}(\vect{S}_{n}^{33})^{-1}\vect{S}_{n}^{32},
  \\
  (\vect{S}_{n}^{-1})^{11}  &=
  \left(\vect{S}_{n}^{11} - \vect{S}_{n}^{12} (\vect{S}_{n}^{22/33})^{-1} \vect{S}_{n}^{21}
    + 2\Re\left[\vect{S}_{n}^{12} (\vect{S}_{n}^{22/33})^{-1} \vect{S}_{n}^{23}(\vect{S}_{n}^{33})^{-1}\vect{S}_{n}^{31}\right]
    - \vect{S}_{n}^{13}(\vect{S}_{n}^{33})^{-1}\vect{S}_{n}^{31}\right. \nonumber \\
  &\qquad\left.
    - \vect{S}_{n}^{13}(\vect{S}_{n}^{33})^{-1} \vect{S}_{n}^{32} (\vect{S}_{n}^{22/33})^{-1} \vect{S}_{n}^{23}
    (\vect{S}_{n}^{33})^{-1}\vect{S}_{n}^{31}\right)^{-1},
\end{align}
\vspace{-15pt}
\begin{align}
  \vect{\rho}_{1} &= \vect{S}_{n}^{12} (\vect{S}_{n}^{22/33})^{-1}
    - \vect{S}_{n}^{13}(\vect{S}_{n}^{33})^{-1}\vect{S}_{n}^{32} (\vect{S}_{n}^{22/33})^{-1},
  \\
  \vect{\rho}_{2} &= -\vect{S}_{n}^{12} (\vect{S}_{n}^{22/33})^{-1} \vect{S}_{n}^{23}(\vect{S}_{n}^{33})^{-1}
    + \vect{S}_{n}^{13}(\vect{S}_{n}^{33})^{-1} + \vect{S}_{n}^{13}(\vect{S}_{n}^{33})^{-1} \vect{S}_{n}^{32}
    (\vect{S}_{n}^{22/33})^{-1} \vect{S}_{n}^{23} (\vect{S}_{n}^{33})^{-1},
  \\
  \vect{\sigma}\ &= \vect{S}_{n}^{23}(\vect{S}_{n}^{33})^{-1}.
\end{align}
\end{widetext}
This reduces the complexity to $\mathcal{O}(N)$.\\

Although this compact representation provides a clearer view of the noise interactions between detectors, the standard multivariate Whittle likelihood may be more suitable for practical implementation. In such representation, the time-domain likelihood is asymptotically decomposed into $N$ independent likelihoods, each dependent on a $3\times 3$ spectral matrix evaluated at the $N$ frequency components.
As a result, the problem of inverting the covariance matrix reduces to the inversion of the $N$ $3\times 3$ spectral matrices, without the necessity to use the analytical expression in Eq.~\eqref{eq: App spectral matrix inversion analytical}.\\
\indent Despite their differences, the two representations are connected through a simple permutation.
In particular, let $\vect{P}$ be the permutation matrix defined as:
\begin{equation}\label{eq: permutation_matrix}
  \vect{P}[i,j] =
  \begin{cases*}
    1 & \text{ if } $i = k + dN$ \text{ and } $j = d + 3k$ \\
    0 & \text{ otherwise}
  \end{cases*}
\end{equation}
with $k = k_\text{low}, \dots, k_\text{high}$ index over the frequency components and $d=1,2,3$ index over the number of detectors.
We have that $\vect{P}$ separates the frequency components of $\vect{S}_n$ as
\begin{equation}\label{eq: permutation spectral matrix}
  \vect{P}^T\vect{S}_n\vect{P} = \vect{S}'_n = \text{diag}\left[\vect{S}_n(f_{k_\text{low}}), \dots, \vect{S}_n(f_{k_\text{high}})\right]
\end{equation}
where each $\vect{S}_n(f_k)$ is the block matrix for the $k\text{-th}$ frequency component used in the standard Whittle approximation:
\begin{equation}
 \vect{S}_n(f_k) = \begin{bmatrix}[1.3]
  S_n^{11}(f_k) & S_n^{12}(f_k) & S_n^{13}(f_k) \\
  S_n^{21}(f_k) & S_n^{22}(f_k) & S_n^{23}(f_k) \\
  S_n^{31}(f_k) & S_n^{32}(f_k) & S_n^{33}(f_k)
 \end{bmatrix}
\end{equation}
As shown, since $\vect{S}'_n$ is block diagonal, its inverse is computed by inverting its $3\times 3$ blocks, which has complexity $\mathcal{O}(N)$~\cite{farebrother2018linear}.
As a final note, applying the inverse permutation on ${(\vect{S}'_n)}^{-1}$ yields from Eq.~\eqref{eq: permutation spectral matrix}:
\begin{equation}
 \vect{P}{(\vect{S}'_n)}^{-1}\vect{P}^T = {(\vect{S}_n)}^{-1}
\end{equation}
This demonstrates that the inverse of the spectral matrix $\vect{S}_n$ retains the same $3\times 3$ block matrix structure, where each block is a $N\times N$ diagonal matrix.

\section{Lower limit for the correlation coefficient}\label{App: Lower limit for the correlation coefficient}
\noindent We provide here a proof for the lower limit of $-0.5$ for the correlation coefficient $\alpha$ used in our simulations.\\

In the assumption that the \ac{PSD} and the \ac{CSD} are the same for all the three \ac{ET} detectors and detector pairs, we have for a certain frequency $f_k$:
\begin{subequations}
\begin{align}
    &\mathbb{E}\left[\tilde{n}_{\ell}(f_k)\tilde{n}_{\ell}^{*}(f_k)\right] = \dfrac{1}{2\Delta f} S_{n}^{\ell \ell}(f_k) = \dfrac{1}{2\Delta f} S_{n}(f_k)\\
    &\mathbb{E}\left[\tilde{n}_{\ell}(f_k)\tilde{n}_{m}^{*}(f_k)\right] = \dfrac{1}{2\Delta f} S_{n}^{\ell m}(f_k) = \dfrac{\alpha_k}{2\Delta f} S_{n}(f_k)
\end{align}
\end{subequations}
with $\ell \neq m$.
By cross-correlating the noise of one detector with the sum of the other two, we obtain
\begin{equation}\label{eq: cross correlation lmr}
    \mathbb{E} \left[
        \tilde{n}_{\ell}(f_k)
        \Big(\tilde{n}_{m}^{*}(f_k) + \tilde{n}_{r}^{*}(f_k)\Big) \right] \hspace{3pt} = \dfrac{\alpha_k}{\Delta f} S_{n}(f_k),
\end{equation}
and from the Cauchy–Schwarz inequality \cite{mukhopadhyay2020probability}, we can write
\begin{widetext}
\begin{subequations}
\begin{align}
    \Big| \hspace{3pt} \mathbb{E} \Big[ \tilde{n}_{\ell}(f_k)
        \Big(\tilde{n}_{m}^{*}(f_k) + \tilde{n}_{r}^{*}(f_k)\Big) \Big] \hspace{3pt} \Big|
    & \leq \sqrt{
        \mathbb{E}\left[\tilde{n}_{\ell}(f_k)\tilde{n}_{\ell}^{*}(f_k)\right] \hspace{3pt}
        \mathbb{E}\Big[\Big(\tilde{n}_{m}(f_k) + \tilde{n}_{r}(f_k)\Big)\Big(\tilde{n}_{m}^{*}(f_k) + \tilde{n}_{r}^{*}(f_k)\Big)\Big]
    } \\
    & = \dfrac{1}{\Delta f} S_{n}(f_k) \sqrt{\dfrac{1+\alpha_k}{2}} \label{eq: Chauchy Schwarz}
\end{align}
\end{subequations}
\end{widetext}
We have therefore, comparing Eqs.~\eqref{eq: cross correlation lmr} -~\eqref{eq: Chauchy Schwarz},
\begin{equation}
    \left| \alpha_k \right| \leq \sqrt{\dfrac{1+\alpha_k}{2}}
\end{equation}
from which $-0.5 \leq \alpha_k \leq 1$.

\end{document}